\documentclass{article}

\usepackage[utf8]{inputenc}

\usepackage[round]{natbib}
\usepackage{color, colortbl}
\usepackage[first=0,last=9]{lcg}

\usepackage[pdftex]{graphicx}

\usepackage{amsmath}
\usepackage[english]{babel}
\usepackage[version=3]{mhchem}
\usepackage{comment}
\usepackage{tikz}

\usepackage{pdfpages}

\definecolor{LightCyan}{rgb}{0.88, 1, 1}
\definecolor{purple}{rgb}{1, .79, 1}
\definecolor{Yellow}{rgb}{1, 1, .6}
\definecolor{Gray}{gray}{0.85}
\definecolor{Almond}{rgb}{0.94, 0.87, 0.8}
\definecolor{antiquebrass}{rgb}{0.8, 0.58, 0.46}
\definecolor{applegreen}{rgb}{0.55, 0.71, 0.0}
\definecolor{babypink}{rgb}{0.96, 0.76, 0.76}
\definecolor{aqua}{rgb}{0.0, 1.0, 1.0}
\usepackage{enumerate}
\usepackage{amsthm}
\usepackage{amsmath, amssymb}

\usepackage[export]{adjustbox}
\usepackage{verbatim}

\usepackage{pgfplots}
\pgfplotsset{compat=1.7}
\usepackage{array}

\usepackage[leftcaption]{sidecap}
\usepackage{pst-node}
\usetikzlibrary{shapes.geometric}%
\usepackage[colorinlistoftodos]{todonotes}
\usepackage[colorlinks=true, allcolors=blue]{hyperref}

\usepackage{setspace}
\usepackage{fullpage}
\usepackage{times}
\usepackage{fancyhdr,graphicx,amsmath,amssymb}
\usepackage[ruled,vlined]{algorithm2e}
\include{pythonlisting}
\usepackage{neuralnetwork}

\makeatletter

\begin{document}

\title{The Bias-Variance Tradeoff of Doubly Robust Estimator with Targeted $L_1$ regularized Neural Networks Predictions}

\author{ 
Mehdi Rostami, Olli Saarela, Michael Escobar 
\\ 
Biostatistics, Dalla Lana School of Public Health, University of Toronto, ON, Canada}

\maketitle

\begin{abstract}

The Doubly Robust (DR) estimation of ATE can be carried out in 2 steps, where in the first step, the treatment and outcome are modeled, and in the second step the predictions are inserted into the DR estimator. The model misspecification in the first step has led researchers to utilize Machine Learning algorithms instead of parametric algorithms. However, existence of strong confounders and/or Instrumental Variables (IVs) can lead the complex ML algorithms to provide perfect predictions for the treatment model which can violate the positivity assumption and elevate the variance of DR estimators. Thus the ML algorithms must be controlled to avoid perfect predictions for the treatment model while still learn the relationship between the confounders and the treatment and outcome.

We use two Neural network architectures and investigate how their hyperparameters should be tuned in the presence of confounders and IVs to achieve a low bias-variance tradeoff for ATE estimators such as DR estimator. Through simulation results, we will provide recommendations as to how NNs can be employed for ATE estimation.

\end{abstract}


\section{Introduction}\label{section1}

There are generally two approaches to address causal inference in observational studies. The first one is to draw population-level causal inference which goes back at least to 1970s \citep{rubin1976multivariate}. The second is to draw conditional causal inference which has received attention recently more \citep{van2007causal, johansson2016learning}. An example of a population-level causal parameter the average treatment effect (ATE),  
\begin{equation}\label{causalriskdiff0}
   \beta_{ATE} = \mathbb{E}[Y^1 - Y^0] = \mathbb{E}\big[\mathbb{E}[Y^1 - Y^0|W]\big].
\end{equation}

The quantity $\mathbb{E}[Y^1 - Y^0|W]$ is referred to as the conditional average treatment effect (CATE) \citep{foster2011subgroup, taddy2016nonparametric, athey2016recursive, li2017causal, wager2018estimation, lu2018estimating, imai2013estimating}. CATE is an individual-level causal parameter which is impossible to estimate accurately unless both potential outcomes are observed for each individual (under parallel worlds!), or $W$ contains all the varying factors that make the causal relationship deterministic. That said, under certain assumptions, the counterfactual loss, the loss due to absence of counterfactual outcome, can be upper bounded \citep{shalit2017estimating}. This article focuses on the estimation of ATE which does not require those assumptions.

To estimate ATE in observational studies, ignoring confounding variables or not taking them into account properly introduces (selection) bias in the estimated causal relationship \citep{angrist2008mostly}. The strong ignorability assumption  \citep{rosenbaum1983central} states that all the confounding variables are contained in the covariates \citep{rubin1976multivariate}. This assumption cannot be verified in real world problems, but allows to draw approximate causal inference.

The Doubly Robust (DR) is one the most common estimators of ATE whose estimation can be carried out in two steps. In step 1, the treatment and outcome are predicted by a statistical or machine learning (ML) algorithm, and in the second step the predictions are inserted into the causal estimator. The ML algorithms in step 1 can capture the linear and nonlinear relationships between the confounders or IVs and the treatment and the outcome.

Through a number of attempts \citep{van2006targeted, belloni2012sparse, belloni2014inference, alaa2017deep, chernozhukov2018double, farrell2018deep}, utilized ML models for the causal parameter estimation. While the ultimate goal of a ML algorithm is to predict the outcome of interest as accurate as possible, it does not optimally serve the main purpose of the causal parameter estimation. In fact, ML algorithms minimize some prediction loss containing the treatment or the observed outcome (and not conterfactual outcome) and without targeting any relevant predictor(s) such as confounding variables \citep{van2011targeted}. In fact, the ML algorithm can successfully learn the linear and non-linear relationships between the confounders and the treatment and outcome, but at the same time, might learn from IVs as well. If there are strong confounders and IVs among the covariates, the predictions of treatments (i.e. the propensity scores) can become extreme (near zero or one) which in turn can make the estimates unstable. This can reduce the bias on the observed data but elevates the variance at the same time. Less complex models, on the other hand, suffer from large bias (under-fitting) but can obtain more stable causal parameter estimation. This conflict between the complexity of model in step 1 and bias-variance tradeoff in step2 motivates to develop ML algorithms for step 1 that provide a compromise between learning from confounders and IVs to entail a balance between the bias and variance of the causal parameter in step 2.

\cite{chernozhukov2018double} investigated the asymptotic normality of orthogonal estimators of ATE (including DR) when two separate ML algorithms (not necessarily NNs) model the treatment and outcome, then the name Double Machine Learning (DML). With the same objective, \cite{farrell2018deep} utilized two separate neural networks (\cite{friedman2001elements} Chapter 11) (we refer to as the double NN or dNN) but without the usage of any regularization other than using the Stochastic Gradient Descent (SGD) for model optimization. SGD does impose some regularization but is insufficient to control the complexity of NN algorithms in practice.

The objective of this research is to experimentally investigate how NN-type methods can be utilized for ATE estimation, and how the hyperparameters can be tuned to achieve the best bias-variance tradeoff for the ATE estimators such as the DR estimator.

The first NN-method is referred to as the joint Neural Network (jNN) algorithm, where we place the treatment and outcome on the output layer of a multi-layer perceptron \citep{friedman2001elements}. This NN architecture is appealing as it models the treatment and outcome simultaneously which might target the relevant covariates that are predictive of both treatment and outcome (or confounder) and can mitigate or ignore the IVs' effects on the predictions. We will investigate if this hypothesis is correct on the experimental data. The second NN-method is dNN utilized by \cite{farrell2018deep}, but we introduce and implement multiple hyperparameters. 

The strategy of targeting specific type of features can be well designed in NN architectures along with the necessary optimization and regularization techniques. Flexible NN structures, optimizations and regularization techniques are easily programmed in the deep learning platforms such as pytorch. 

To show the effectiveness of jNN and dNN, a thorough simulation study is performed and these methods are compared in terms of the number of confounders and IVs that are captured in each scenario, the prediction measures, and the bias and variance of causal estimators.

To investigate whether our network targets confounders rather than IVs and also dampen the impact of strong confounders on the propensity scores, we calculate the bias-variance tradeoff of causal estimators (i.e. minimal MSE) utilizing the NN predictions; Low bias means the model has mildly learned from confounders and other types of covariates for the outcome, and low variance means model has ignored IVs and has dampened strong confounders in the treatment model.

The organization of this paper is as follows. In section \ref{setting} we define the problem setting and the causal parameter to be estimated, and its estimator. In section \ref{methods} we introduce the NN-type methods, their loss functions, and hyperparameters. Section \ref{ate} provides a quick review of the ATE estimators. In section \ref{simulations} our simulation scenarios are stated along with their results. We conclude the paper in section \ref{discussion} with some discussion on the results and future work.


\subsection{Notation}

\begin{figure}[!htbp]
\centering
\includegraphics[ scale=0.6]{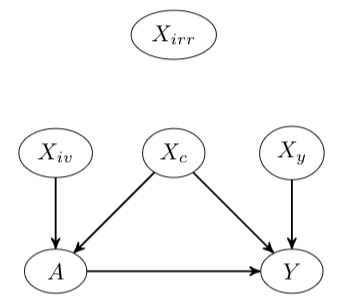}
\caption{The causal relationship between $A$ and $y$ in the presence of other factors in an observational setting. }
\label{firstdiagram}
\end{figure}

Let data $\mathbf{O}=(O_1, O_2, ..., O_n)$ be generated by a data generating process $F$, where $O_i$ is a finite dimensional vector $O_i=(Y_i, A_i, W_i)$, with $\mathbf{W}=(X_c, X_y, X_{iv}, X_{irr})$. Diagram \ref{firstdiagram} visualizes how the adjusting factors $X_j$'s are related to $A$ and $Y$ which assumes $A = f_1(X_c, X_{iv})+\epsilon_1$, $y = f_2(A, X_c, X_{y})+\epsilon_2$, for some random functions $f_1, f_2$. We denote the prediction function of observed outcome given covariates in the treated group $Q(1, W)=\mathbb{E}[Y|A=1, W]$, and that in the untreated group $Q(0, W)=\mathbb{E}[Y|A=0, W]$, and the propensity score as $g(W)=\mathbb{E}[A|W]$. The symbol $\ \hat{}\ $ on the population-level quantities indicates the corresponding finite sample estimator.

\vspace{.5cm}

    

\subsection{Problem Setup and Assumptions}\label{setting}

The fundamental problem of causal inference states that individual-level causality cannot be exactly determined since each person can experience only one value of $A$. Thus, it is customary to only estimate a population-level causal parameter, in this research Average Treatment Effect (ATE) \eqref{causalriskdiff0}.

For identifiablity of the parameter, the following assumptions must hold true. The first assumption is the Conditional Independence, or Unconfoundedness stating that, given the confoudners, the potential outcomes are independent of the treatment assignments ($Y^0, Y^1 \perp A| W$). The second assumption is Positivity which entails that the assignment of treatment groups is not deterministic ($0 < Pr(A = 1|W) < 1$). The third assumption is Consistency which states that the observed outcomes equal their corresponding potential outcomes ($Y^{A} = y$). There are other modeling assumptions made such as time order (i.e. the covariates $W$ are measured before the treatment), IID subjects, and a linear causal effect. 

\section{Prediction Models}\label{methods}

Neural Networks (NNs) are complex nonparametric models that estimate the underlying relationship between inputs and the outcome. The objective in causal inference, however, is not necessarily to leverage the maximum prediction strength of NNs and in fact, the NN architecture should be designed and tuned so that it pays more attention to the confounders. 

The most important requirement of ML models such as NNs in causal inference is that although the outcome prediction model should minimize the corresponding loss (fit to get the best outcome prediction possible), given all of the covariates, the loss function associated with the propensity score model should not necessarily be minimized. Ideally, the instrumental variables or strong confounders which can give extreme probability values (near zero or one) should be controlled when minimizing the loss. This helps prevent the elevation of the variance of the causal estimator (i.e. prevent the violation/near violation of the positivity assumption \citealp{van2011targeted}). In summary, the prediction models should be strong enough to learn the linear and non-linear relationships between the confounders and treatment, but should not provide perfect predictions. We hypothesize that the employed NNs methods with the regularization techniques have the property of ignoring or damping strong confouders and/or instrumental variables.

\begin{figure}
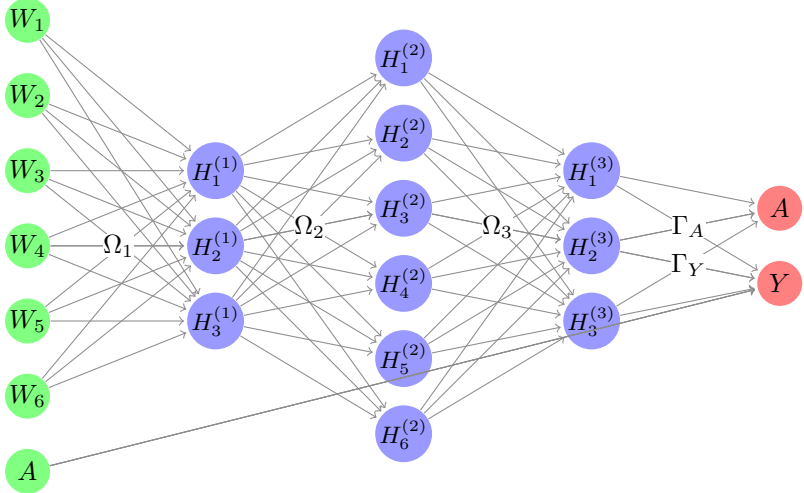

\begin{neuralnetwork}[height=7]
    \newcommand\nn@nodeindex{0}
    \newcommand{\y}[2]{$\hat{y}_#2$}
    
    \newcommand{\hfirst}[2]{\small $H^{(1)}_#2$}
    \newcommand{\hsecond}[2]{\small $H^{(2)}_#2$}
    \newcommand{\hthird}[2]{\small $H^{(3)}_#2$}

    \newcommand{\x}[2]{
        \ifnum \nn@nodeindex=7
            $A$
        \else
            $W_#2$
        \fi}
        
    \newcommand{\outlayer}[2]{
        \ifnum \nn@nodeindex=1
            $A$
        \else
            $Y$
        \fi}
    \newcommand{\w}[4] {$\Omega_#3$}
    \newcommand{\gy}[4] {$\Gamma_Y$}
    \newcommand{\gA}[4] {$\Gamma_A$}
    
    \inputlayer[count=7, bias=false, text=\x]

    \hiddenlayer[count=3, bias=false, text=\hfirst] 

    \hiddenlayer[count=6, bias=false, text=\hsecond] \linklayers
    
    \hiddenlayer[count=3, bias=false, text=\hthird] \linklayers
    
    \outputlayer[count=2, text=\outlayer] \linklayers
    
    \foreach \n in {1,...,6}{ 
        \foreach \m in {1,..., 3}{ 
            \link[from layer=0, from node=\n, to layer=1, to node=\m] 
        } 
    }

    \link[style={}, from layer=0, from node=7, to layer=4, to node=2] 
    \link[style={}, from layer=0, from node=7, to layer=4, to node=2] 
    \link[style={}, from layer=0, from node=7, to layer=4, to node=2] 

    \link[style={}, label=\w, from layer=0, from node=4, to layer=1, to node=2] 
    \link[style={}, label=\w, from layer=1, from node=2, to layer=2, to node=3] 
    \link[style={}, label=\w, from layer=2, from node=3, to layer=3, to node=2] 
    
    \link[style={}, label=\gA, from layer=3, from node=2, to layer=4, to node=1] 
    \link[style={}, label=\gy, from layer=3, from node=2, to layer=4, to node=2] 
\end{neuralnetwork}
\caption{An IV Mitigator Neural Net architecture that incorporates linear effect of the treatment on the outcome, and the nonlinear relationship between the covariates and the treatment assignment and the outcome, all three tasks at the same time. }
\label{jNNarch}
\end{figure}

\subsection{Joint Neural Nettork}\label{jNNNN}

The joint Neural Nettork (jNN) architecture is a combination of multiple ideas (see below \ref{multitasksec} - \ref{skipconn}) for causal parameter estimation purposes mentioned above.

The jNN models are:

\begin{equation}\label{jNNeq}
\left[ \begin{array}{c} \mathbb{E}[Y|A, W]\\ \mathbb{E}[A|W]  \end{array} \right] = \begin{bmatrix} \alpha_0 + \beta A + \mathbf{W}\alpha + \mathbf{H}\mathbf{\Gamma}_Y \\ g(\gamma_0 + \mathbf{W}\gamma + \mathbf{H}\mathbf{\Gamma}_A) \end{bmatrix}
\end{equation}
where $\mathbf{H}=f(f(...(f(\mathbf{W}\mathbf{\Omega}_1)\mathbf{\Omega}_2)...)\mathbf{\Omega}_L)$ is the last hidden layer matrix which is a non-linear representation of the inputs ($L$ is the number of hidden layers), $g$ is the logistic link function, and $\mathbf{\Gamma}_A$ and $\mathbf{\Gamma}_Y$ are the parameters that regress $\mathbf{H}$ to the log-odds of the treatment assignment or to the outcome in the output layer. The large square brackets around the equations above is meant to emphasize that both treatment and outcome models are trained jointly.

Let $\mathcal{P}$ be the set of shared parameters of the non-linear part of the network. The jNN architecture minimizes a multi-task loss \ref{multitasksec} to estimate the networks parameters:
\vspace{-.2cm}
\begin{multline}\label{jNNloss}
    L(\mathcal{P}, \beta, \alpha) = a\sum_{i=1}^n \Big[y_i - \alpha' - \beta A_i -\mathbf{W}_i\alpha - H_i^T\mathbf{\Gamma}_Y \Big]^2 + \\
    b \sum_{i=1}^n \Big[A_i \log\Big(g\big(H_i^T\mathbf{\Gamma}_A\big)\Big) + (1-A_i) \log\Big(1-g\big(H_i^T\mathbf{\Gamma}_A\big)\Big)\Big] +\\ C_{L_1}\sum_{\omega \in \mathcal{P}} |\omega| + C_{L_1TG}\big(\sum_{\omega \in \mathbf{\Gamma}_A} |\omega|+ \sum_{\omega \in \mathbf{\Omega}_1} |\omega|\big)
\end{multline}
where $a, b, , C_{L_1}, C_{L_1TG}$ are hyperparameters, that can be set before training or be determined by Cross-Validation, that can convey the training to pay more attention to one part of the output layer. The jNN can have an arbitrary number of hidden layers, or the width of the network ($\mathcal{HL}$) is another hyperparameter. For a 3-layer network, $\mathcal{HL}=[l_1,l_2, ...,l_{h}]$, where $l_j$ is the number neurons in layer $j$, $j=1, 2, ..., h$. $\mathcal{P}=\{\omega \in \Omega_1\cup\Omega_2\cup\Omega_3\cup\Gamma_Y\cup\Gamma_A\}$, are the connection parameters in the nonlinear part of the network, with $\Omega$'s being shared for the two outcome and propensity models. Noted that the number of parameters with $L_1$ regularization (third term on \eqref{jNNloss}) is $|\mathcal{P}|=(p+1)\times l_1 + (l_1+1)\times l_2 + ... + (l_{h-1}+1)\times l_h + (l_h+1)\times 2$, including the intercepts in each layer.

The following subsections list the potential benefits and the rationale behind the proposed network (equations \eqref{jNNeq} and \eqref{jNNloss}).

\subsection{Bivariate Prediction, Parameters Sharing, and Multi-task Learning}\label{multitasksec}

One of the main components of the jNN architecture is that both treatment and outcome are placed and modeled in the output layer simultaneously. The hypothesis here is that the network learns to get information from the inputs that predict both treatment and outcome, i.e. the confounders. This bivariate structure is intertwined with a multi-task learning or optimization. \citet{ruder2017overview} reviews the multi-tasking in machine learning and lists its benefits such as implicit data augmentation, regularization, attention focusing, Eavesdropping and Representation bias. \citet{caruana1995learning} showed that overfitting declines by adding more nodes to the output layer as compared to modeling each output separately \citep{baxter1997bayesian}. The multi-task is used when more than one output is used. Multi-task learning is common in the field of Artificial Intelligence and Computer Vision, for example, for the object detection task where the neural network predicts the coordinates of the box around objects and also classifies the object(s) inside the box (see for example \cite{yolo2016}). Multi-task learning is used in jNN in order to investigate if the model pays more attention to the confounder than other types of inputs.

\subsection{Regularization}

The jNN will be resistant to overfitting by adding regularization to the network. Preliminary simulations revealed that $L_2$, and the Dropout \citep{goodfellow2016deep} regularization techniques do not result in satisfactory causal effect estimation, and the inherent regularization in the Stochastic Gradient Descent \citep{goodfellow2016deep} is also insufficient, while $L_1$ regularization is effective. 

The $L_1$ regularization, third summation in \eqref{jNNloss}, shrinks the magnitude of the parameter estimates of the non-linear part of the architecture which, in effect, limits the influence of $X_{irr}$ and $X_{iv}$, $X_y$, and $X_c$ on both treatment and the outcome. The motivation behind the  $L_1$ regularization is to avoid overfitting for better generalization.

The ideal situation for causal parameter estimation is to damp the instrumental variables and learn from confounders and y-predictors only. Henceforth another version of the $L_1$ regularization is introduced here, referred to as the targeted $L_1$ regularization, or $L_1TG$, to potentially reduce the impact of instrumental variables on the outcome and more importantly on the propensity scores. The motivation is that by introducing shrinkage on the connections between the last hidden layer and the treatment, the neural network is trained to learn more about confounders than IVs in the last hidden layer as the outcome model is free to learn as much as possible from confounders. The caveat here might be that if the last hidden layer is large enough, some of the neurons can learn confounders while other learn from IVs, thus motivating to consider limiting the number of neurons in the last hidden layer. These hypotheses and ideas are considered in the simulation studies.

\subsection{Linear Effects and Skip Connections}\label{skipconn}

The terms $\beta A + \mathbf{W}\alpha$ and $\mathbf{W}\gamma$ in \eqref{jNNeq} are responsible for potential linear effects. Theoretically, the non-linear parts of the NNs can estimate linear effects, but it is preferable to use linear terms if the relationship between the some of the inputs and the outcome/treatment are linear for more accurate linear effect estimation. The benefit of including linear terms in the equations has been verified in our preliminary simulation studies.

These linear terms are referred to the skip-connections in ML literature \citep{he2016deep} which connect some layers to two or more layers forward. In ML literature, they are primarily used in very deep neural networks to facilitate optimizations. But they are used in jNN to model the linear effects directly. More specifically, skip connections connect the covariates to both treatment and outcome in the output layers and connect the treatment in the input layer to the outcome in the output layer. The latter skip connection is shown in Figure \ref{jNNarch}. It should be noted that this skip connection in particular is independent of the treatment in the output layer to avoid perfect prediction of the propensity scores.

\subsection{Double Neural Networks}\label{dNNNN}

In order to study the significance of the proposed method through simulations, we compare jNN with the double Neural Networks (dNN) \citep{chernozhukov2016double} method. dNN is generally referred to the strategy of modeling the treatment and outcome separately utilizing two different models:

\begin{equation} \label{eq1dNN}
\begin{split}
        \mathbb{E}[Y|A, W] &= \beta_0 + \beta A + \mathbf{W}\alpha + \mathbf{H}\mathbf{\Gamma}_Y\\
        \mathbb{E}[A|W] &= \beta'_0 + \mathbf{W'}\alpha' + \mathbf{H'}\mathbf{\Gamma}_A,
\end{split}
\end{equation}
where two separate neural nets model $y$ and $A$ (no parameter sharing). In this paper, the dNN algorithm refers to two neural networks to model the treatment and outcome separately. To make the two jNN and dNN models comparable, we let the NN architectures to be as similar as possible in terms of skip connections and regularization techniques. The loss functions in dNN to be optimized are:

\begin{equation} \label{eq1lossdNN}
\begin{split}
        L_y(\mathcal{P}_y, \beta, \alpha) = & \sum_{i=1}^n \Big[y_i - \alpha' - \beta A_i -\mathbf{W}_i\alpha - H_i^T\mathbf{\Gamma}_Y \Big]^2 + C'_{L_1}\sum_{\omega \in \mathcal{P}} |\omega|,\\
        L_A(\mathcal{P}_A) = & \sum_{i=1}^n \Big[A_i \log\Big(g\big(H_i^T\mathbf{\Gamma}_A\big)\Big) + (1-A_i) \log\Big(1-g\big(H_i^T\mathbf{\Gamma}_A\big)\Big)\Big] + C''_{L_1}\sum_{\omega \in \mathcal{P}} |\omega| +\\ &C_{L_1TG}\big(\sum_{\omega \in \mathbf{\Gamma}_A} |\omega| + \sum_{\omega \in \mathbf{\Omega}_1} |\omega|\big),
\end{split}
\end{equation}

\section{ATE Estimation}\label{ate}

The Causal Parameter Estimation algorithm is a two stage process. The regression functions $\mathbb{E}[A |W], \ \mathbb{E}[Y |A=1, W]),\  \mathbb{E}[Y |A=0, W]$ are estimated using the ML algorithms such as jNN or dNN in step 1. And in step 2, the predictions are inserted into the causal estimators such as \eqref{allests}, below.

\subsection{ATE Estimators}

There is a wealth of literature on how to estimate the ATE and there are various versions of estimators including the naive Average Treatment Effect (nATE), Doubly Robust (DR), Normalized Doubly Robust (nDR), Robinson II:

\begin{equation} \label{allests}
\begin{split}
        \hat{\beta}_{Rob1} &= \Big(\sum_{i=1}^n \hat{V}_i^2\Big)^{-1}\sum_{i=1}^n \hat{V}_i\hat{e}_i,\\ 
        \hat{\beta}_{Rob2} &= \Big(\sum_{i=1}^n \hat{V}_iA_i\Big)^{-1}\sum_{i=1}^n \hat{V}_i\hat{e}_i,\\ \\
        \hat{\beta}_{DR} &= 
   \frac{1}{n}\sum_{i=1}^n \Big(\frac{A_i(y_i-\hat{Q}^1_i)}{\hat{g}_i} - \frac{(1-A_i)(y_i-\hat{Q}^0_i)}{1-\hat{g}_i}\Big) + \frac{1}{n}\sum_{i=1}^n \hat{Q}^1_i - \hat{Q}^0_i,\\ 
        \hat{\beta}_{nDR} &= \sum_{i=1}^n \Big(\frac{A_i(y_i-\hat{Q}^1_i)w_i^{(1)}}{\sum_{j=1}^n A_jw_j^{(1)}} - \frac{(1-A_i)(y_i-\hat{Q}^0_i)w_i^{(0)}}{\sum_{j=1}^n (1-A_j)w_j^{(0)}}\Big) + \frac{1}{n}\sum_{i=1}^n \hat{Q}^1_i - \hat{Q}^0_i.
\end{split}
\end{equation}
where $\hat{Q}^k_i = \hat{Q}(k,W_i) = \hat{\mathbb{E}}[Y_i|A_i=k, W_i]$ and $\hat{g}_i = \hat{\mathbb{E}}[A_i |W_i]$,  $w^{(1)}_k = \frac{1}{\hat{g}_k}$, $w^{(0)}_k = \frac{1}{1-\hat{g}_k}$, $e_i = y_i - \hat{\mathbb{E}}[Y_i|W_i]$, $V_i = A_i - \hat{g}_i$, and $A_1$ is the treatment group with size $n_1$ and  $A_0$ is the treatment group with size $n_1$. 

In the second step of estimation procedure, the predictions of the treatment $\hat{g}(W_i)$ (i.e. propensity score, PS) and/or the outcome $\hat{\mathbb{E}}[Y_i|A_i=k, W_i]$, $k=0,1$, can be inserted in these estimators \eqref{allests}. Generalized Linear Models (GLM), any relevant Machine Learning algorithm such as tree-based algorithms and their ensemble \citep{friedman2001elements}, SuperLearner \citep{van2007super}, or Neural Network-based models (such as ours) can be applied as  prediction models for the first step prediction task. We will jNN and dNN in this article.

\section{Simulations}\label{simulations}

A simulation study (with 100 iterations) was performed to compare the prediction methods jNN, and dNN by inserting their predictions in the NDR (causal) estimators \eqref{allests}. There are a total of 8 scenarios according to the size of the data (i.e. the number of subjects and number of covariates), and the confounding and instrumental variables strengths. We fixed the sample sizes to be  $n = 750$ and $n = 7500$ , with the number of covariates $p = 32$ and $p = 300$, respectively. The four sets of covariates had the same sizes $\#X_c  = \#X_{iv} = \#X_y = \#X_{irr} = 10, 50$ and independent from each other were drawn from the Multivariate Normal (MVN) Distribution as $X \sim \mathcal{N}(\mathbf{0}, \Sigma)$, with $\Sigma_{kj} = \rho^{j-k}$ and $\rho=0.5$. The models to generate the treatment assignment and outcome were specified as
\begin{equation} \label{generatedata}
\begin{split}
    A &\sim Ber(\frac{1}{1+e^{-\eta}}), \text{with} \ \eta = f_a(X_c)\gamma_c + g_a(X_{iv})\gamma_{iv},\\
    y &= 3 + A + f_y(X_c)\gamma'_c + g_y(X_y)\gamma_y + \epsilon,
\end{split}
\end{equation}

and $\beta = 1$. The functions $f_a, g_a, f_y, g_y$ select 20\% of the columns and apply interactions and non-linear functions listed below \eqref{nonlinearfs}. The strength of instrumental variable and confounding effects were chosen as $\gamma_c, \gamma_c', \gamma_y \sim Unif(r_1, r_2)$ where  $(r_1=r_2=0.1)$ or $(r_1=0.1, r_2=1)$, and $\gamma_{iv} \sim Unif(r_3, r_4)$ where $(r_3=r_4=0.1)$ or $(r_3=0.1, r_4=1)$.

The non-linearities are randomly selected among the following functions:
\begin{equation} \label{nonlinearfs}
\begin{split}
    & l(x_1, x_2) = e^{\frac{x_1x_2}{2}}\\
    & l(x_1, x_2) = \frac{x_1}{1+e^{x_2}}\\
    & l(x_1, x_2) = \big(\frac{x_1x_2}{10}+2\big)^3\\
    & l(x_1, x_2) = \big(x_1+x_2+3\big)^2\\
    & l(x_1, x_2) = g(x_1) \times h(x_2)
\end{split}
\end{equation}
where $g(x) = -2 I(x \leq -1) - I(-1 \leq x \leq 0) + I(0 \leq x \leq 2)+ 3 I(x \geq 2)$, and $h(x) = -5 I(x \leq 0) - 2 I(0 \leq x \leq 1) + 3 I(x \geq 1)$, or $g(x) = I(x \geq 0)$, and $h(x) = I(x \geq 1)$.

In order to find the best set of hyperparameter values for the NN architectures, we ran an initial series of simulations to find the best set of hyperparameters for all scenarios, presented here. The networks' activation function is Rectified Linear Unit (ReLU), with 3 hidden layers as large as the input size (p), with $L_1$ regularization and batch size equal to $3*p$ and 200 epochs. The Adaptive Moment Estimation (Adam) optimizer \citep{kingma2014adam} with learning rate 0.01 and momentum 0.95 were used to estimate the network's parameters, including the causal parameter (ATE).

As in practice the RMSE and covariate types are unknown, prediction measures of the outcome and treatment should be used to choose the best model in a K-fold cross-validation. $R^2$ and $AUC$ each provide insight about the outcome and treatment models, respectively, but in our framework, both models should be satisfactory. To measure the goodness of the prediction models (jNN and dNN) for causal inference purposes, we define and utilize a statistic which is a compromise (geometric average) between $R^2$ and $AUC$, here referred to as $geo$,

\begin{equation}\label{r2auc}
    geo = \sqrt[3]{R^2 \times D \times (1 - D)},
\end{equation}
where $D = 2(AUC-0.5)$, the Somers' D index. This measure was not utilized in the optimization process, and is rather introduced here to observe if the compromise between $R^2$ and $AUC$ agrees with the models that capture more confounders than IVs.

\subsection{Selected Covariate types}\label{seletedIVs}

In order to identify which types of covariates (confounders, IVs, y-predictors, and irrelevant covariates) the prediction methods have learned from, we calculate the association between the inputs and the predicted values ($\hat{\mathbb{E}}[Y |A, W]$ and $\hat{\mathbb{E}}[A |W]$), and after sorting the inputs (from large to small values) based on the association values, we count the number of different types of covariates within top 15 inputs. 

The association between two variables here is estimated using the distance correlation statistic \citep{szekely2007measuring} whose zero values entail independence and non-zero values entail statistical dependence between the two variables.

Figures \ref{simresults1}-\ref{simresults4} present the overall comparison of different hyperparameter scenarios of jNN and dNN predictions in terms of five different measures, respectively: 1) The average number of captured confounders/IVs/y-predictors, 2) Average Root Mean Square Error (RMSE) of causal estimators, 3) Average $R^2$, $AUC$ and their mixture measure $geo$, 4) Bias, 5) MC standard deviation of nDR. The bootstrap confidence intervals for the bias, standard deviation and RMSE are calculated to capture significant differences between the simulation scenarios.
The x-axis includes 16 hyperparameter settings, and as a general rule here, models in the left are most complex (less regularization and wider neural nets) and in the right are least complex. Noted that $L_1TG$ regularization is only targeted at the treatment model.

Overall, the trends favor the idea that more complex treatment models capture larger number of IVs, have larger RMSE and have larger $AUC$ (smaller $geo$.) That is, more complex models are less favorable.

The Figures \ref{simresults1} and \ref{simresults2} show how the complexity of both dNN and jNN (x-axis) impact the number of captured covariate types (i.e. confounders/IVs/y-predictors) (top graph), RMSE (middle graph) and prediction measures (bottom graph). In particular, it is seen that the outcome model remains unchanged throughout the scenarios. This is predictable for dNN model, but it turns out to be the same for the jNN model. That is, the targeted regularization in jNN does not impact the performance of the outcome model. The $AUC$, on the other hand, declines with higher values of $C_{L_1TG}$, and is almost always smaller for jNN than for dNN.

The RMSE of jNN is larger than that of dNN for models with the least amount of regularization (the scenarios in the left). With decreasing the complexity of the model, the RMSE of both jNN and dNN decline. The jNN outperforms dNN in almost all of the hyperparameter settings in case of $n=750$, but fluctuates in case of $n=7500$. Further, the impact of width of architectures ($HL$) changes based on $C_{L_1}$ regularization: wider architectures
($HL=[p, p, p]$, $p$: number of covariates) with large $C_{L_1}$ outperforms other combinations of these two hyperparameters. This observation is more clear for smaller size data, and for dNN model. But overall, the less complex model, the higher RMSE. In the best scenarios, the MSE confidence intervals of IMV model are below those of dNN, illustrating a small preference of jNN over dNN in terms of MSE. Comparing the three hyperparameters, $C_{L_1TG}$ is most effective, and zero values of this hyperparameter results in very large MSEs for both dNN and jNN.

The $geo$, \eqref{r2auc}, is a compromise measure of both $R^2$ and $AUC$, providing insight about both models at the same time. From Figures \ref{simresults1} and \ref{simresults2}, it is observed that both jNN and dNN models have roughly the same values across hyperparameter settings and for both data sizes ($n=7500$, and $n=750$). The hyperparameter sets where the both $AUC$ and $geo$ coincide correspond to best RMSE. 

Figures \ref{simresults3} and \ref{simresults4} illustrate the bias and standard deviation of the causal estimators. As expected and mentioned in the Section \ref{section1}, the models that do not damp IVs suffer from large bias and standard deviation. The bias and standard deviation have opposite behaviour in different scenarios, such that scenarios that produce larger standard deviation, results in small bias, and vice versa. The fluctuations of the bias-variance across hyperparameter settings are larger for $n=750$ case vs. $n=7500$. For small sample $n=750$, the best scenario for jNN is $HL=[32,32,32],C_{L_1}=0.1,C_{L_1TG}=0.7$ where both bias and standard deviation of jNN are small in the same direction. For the large sample $n=7500$, however, the best scenario for jNN is $HL=[30,300,30],C_{L_1}=0.01,C_{L_1TG}=0.7$ with a similar behaviour. The best scenarios for dNN are slightly different. For small sample  $HL=[32,32,32],C_{L_1}=0.1,C_{L_1TG}=0.7$ and for the large sample $HL=[30,300,30],C_{L_1}=0.01,C_{L_1TG}=0.7$ are most favorable.

\section{Discussion}\label{discussion}



\begin{figure}
\centering
\caption{The comparison of captured number of confounders, IVs and y-predictors, RMSE of nDR and its bootstrap 95\% confidence interval, and prediction measures $R^2$, $AUC$ and $geo$ (geometric mean of $R^2$, $AUC$) for different hyperparameter settings and where the predictions come from jNN or dNN models. (n=750, p=32)}
\includegraphics[clip, trim=0cm 1cm 0.5cm 1cm, scale=0.9, width=.9\textwidth]{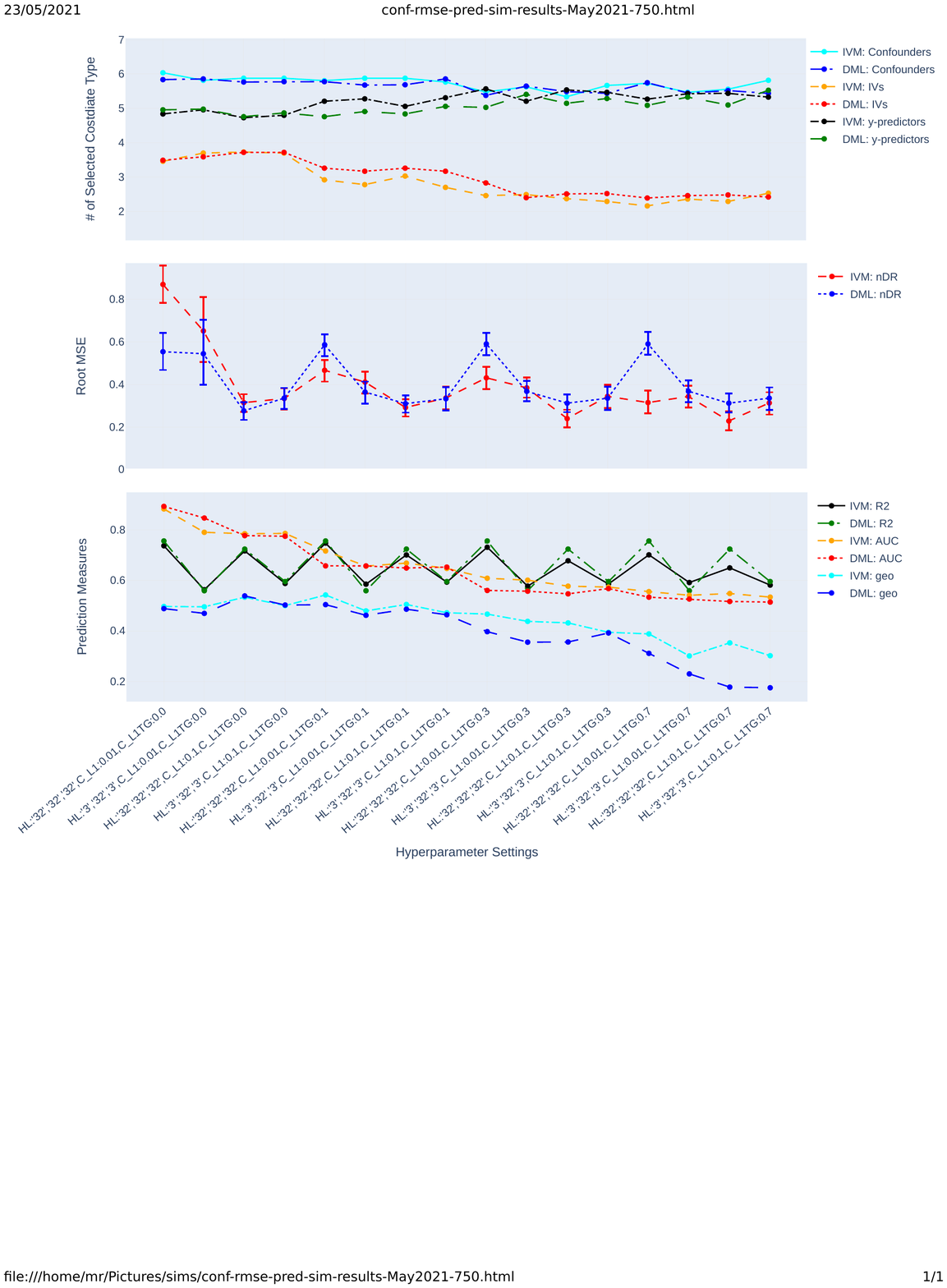}
\label{simresults1}
\end{figure}

\begin{figure}
\centering
\caption{The comparison of captured number of confounders, IVs and y-predictors, RMSE of nDR and its bootstrap 95\% confidence interval, and prediction measures $R^2$, $AUC$ and $geo$ (geometric mean of $R^2$, $AUC$) for different hyperparameter settings and where the predictions come from jNN or dNN models. (n=7500, p=300)}
\includegraphics[clip, trim=0cm 1cm 0.5cm 1cm, scale=0.9, width=.9\textwidth]{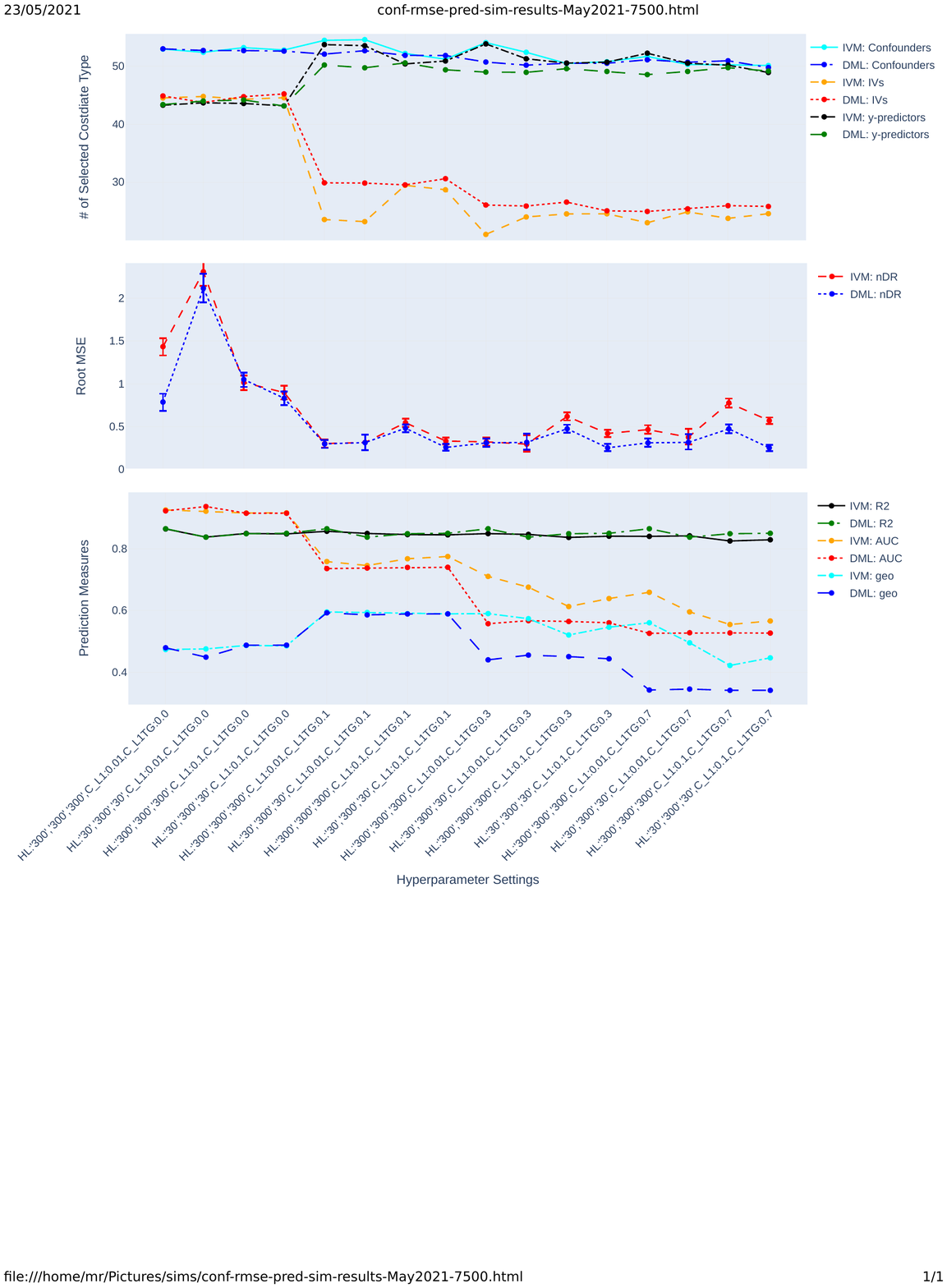}
\label{simresults2}
\end{figure}

\begin{figure}
\centering
\caption{The bias and standard deviation of nDR and their bootstrap 95\% confidence intervals for different hyperparameter settings where the predictions come from jNN or dNN models. (n=750, p=32)}
\includegraphics[clip, trim=0cm 1cm 0.5cm 1cm, scale=.9,width=.9\textwidth]{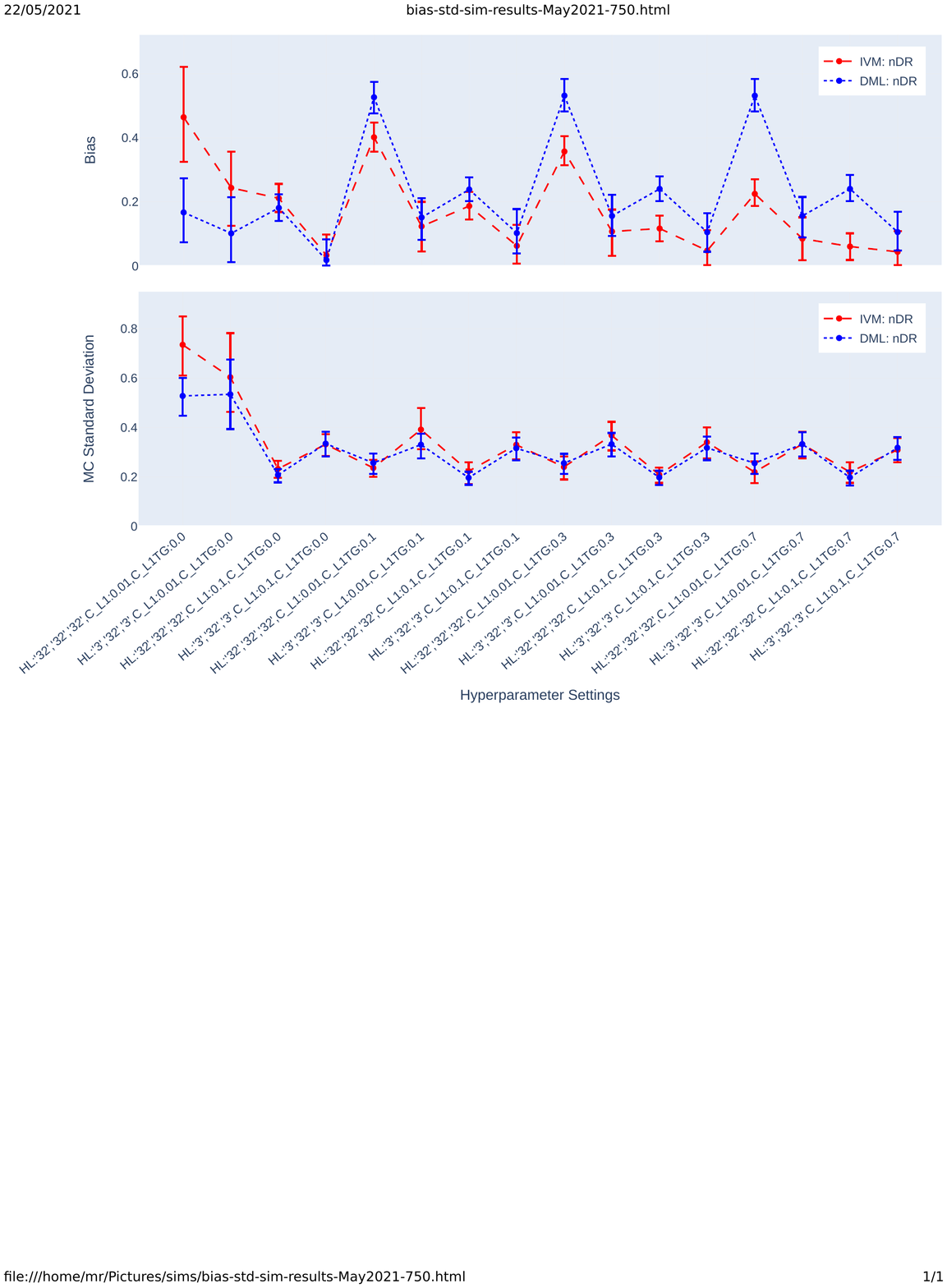}
\label{simresults3}
\end{figure}

\begin{figure}
\centering
\caption{The comparison of bias, Montne Carlo standard deviation and their bootstrap 95\% confidence intervals of nDR, for different hyperparameter settings and the predictions come from jNN or dNN models. (n=7500, p=300)}
\includegraphics[clip, trim=0cm 1cm 0.5cm 1cm, scale=0.9, width=.9\textwidth]{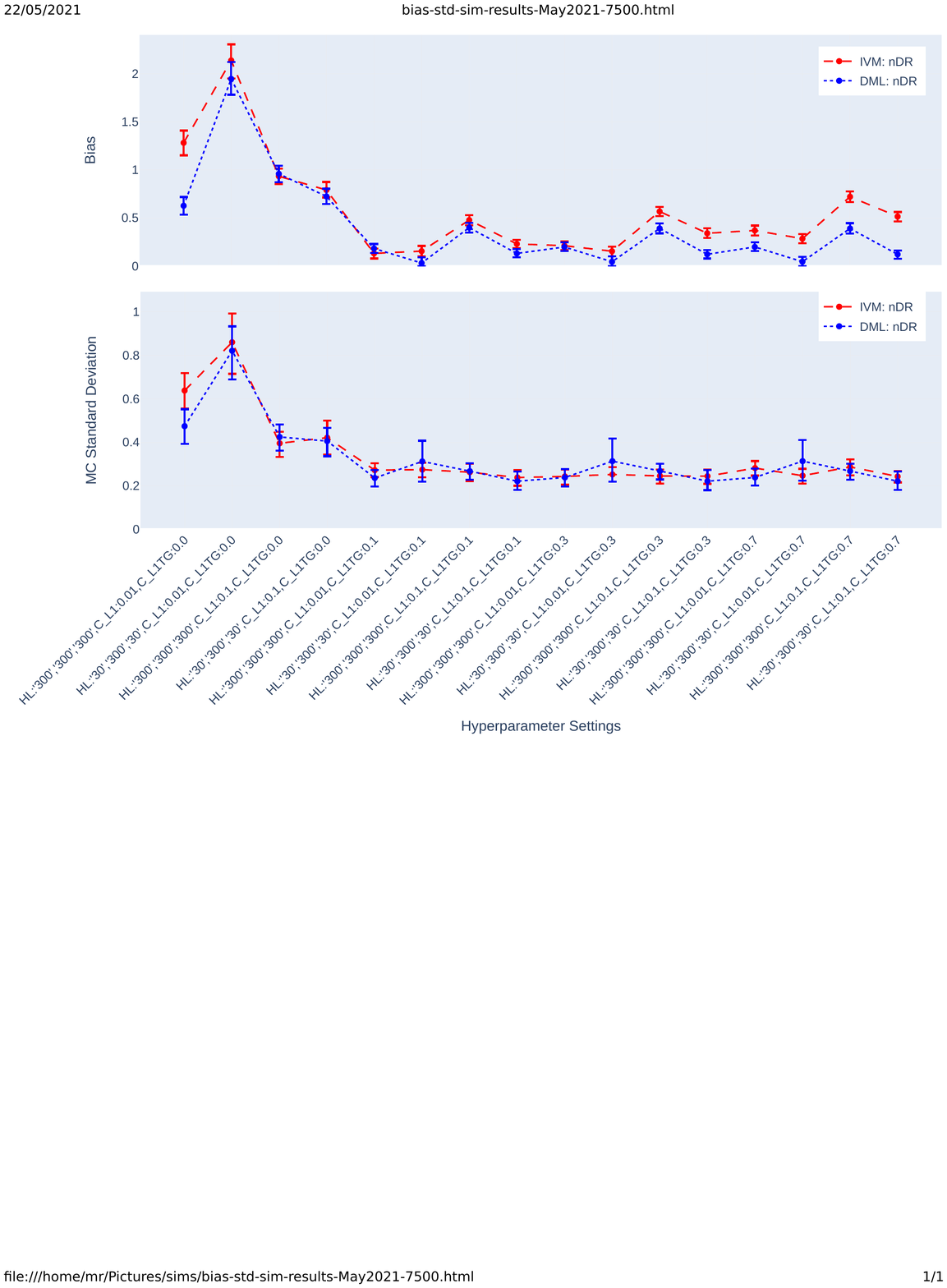}
\label{simresults4}
\end{figure}

In this paper, we have studied how neural networks can be utilized in causal estimation. We have considered a general scenario that four types of covariates exist in the dataset, confounders, IVs, y-predictors and irrelevant covariates. We have observed that $L_1$ regularization especially the ones that targets the treatment model ($L_1TG$) is an influential hyperparameter for achieving the best bias-variance trade-off for the normalized Doubly Robust (nDR) estimator, that is lower Root Mean Saure Error (RMSE). And, the number of neuron in the first and last layer of the network becomes irrelevant as long as the value of $L_1TG$ is sufficient. Further, we have observed that in the hyperparameter scenarios that the IVs effects are controlled, the estimation is less biased and more stable. Thus the regularization techniques have been successful in damping the IVs and preventing perfect prediction in the treatment model. Figures \ref{simresults1}-\ref{simresults4} illustrate that jNN is overall more stable and has a smaller RMSE in the small sample dataset scenario as compared to dNN.

There are limitations due to the assumptions and simulation scenarios and some questions are left to future to be answered. The outcome here was assumed to be continuous, and the treatment to be binary. We also did not cover scenarios where the outcome is a heavy tail, or binary or the treatment is extremely rare. Also, the ratio of dimension to size of the data was considered to be fairly small ($p<<n$), and we have not studied the case where $n<p$. Furthermore, we did not study the asymptotic behaviour of nDR when jNN or dNN predictions are used.

Further, we only utilized neural networks to learn the underlying relationships between the covariates and the outcome and treatment. Other Machine Learning algorithms such as tree-based models, and more specifically Gradient Boosting Machines (GBM) \citep{friedman2001greedy} can be alternatively used to learn these non-linear relationships. We believe GBM can effectively be programmed in pytorch using its automatic differentiation capability. This is postponed to a future article.

\clearpage
\bibliography{ref}

\begin{thebibliography}{32}
\providecommand{\natexlab}[1]{#1}
\providecommand{\url}[1]{\texttt{#1}}
\expandafter\ifx\csname urlstyle\endcsname\relax
  \providecommand{\doi}[1]{doi: #1}\else
  \providecommand{\doi}{doi: \begingroup \urlstyle{rm}\Url}\fi

\bibitem[Alaa et~al.(2017)Alaa, Weisz, and Van Der~Schaar]{alaa2017deep}
A.~M. Alaa, M.~Weisz, and M.~Van Der~Schaar.
\newblock Deep counterfactual networks with propensity-dropout.
\newblock \emph{arXiv preprint arXiv:1706.05966}, 2017.

\bibitem[Angrist and Pischke(2008)]{angrist2008mostly}
J.~D. Angrist and J.-S. Pischke.
\newblock \emph{Mostly harmless econometrics: An empiricist's companion}.
\newblock Princeton university press, 2008.

\bibitem[Athey and Imbens(2016)]{athey2016recursive}
S.~Athey and G.~Imbens.
\newblock Recursive partitioning for heterogeneous causal effects.
\newblock \emph{Proceedings of the National Academy of Sciences}, 113\penalty0
  (27):\penalty0 7353--7360, 2016.

\bibitem[Baxter(1997)]{baxter1997bayesian}
J.~Baxter.
\newblock A bayesian/information theoretic model of learning to learn via
  multiple task sampling.
\newblock \emph{Machine learning}, 28\penalty0 (1):\penalty0 7--39, 1997.

\bibitem[Belloni et~al.(2012)Belloni, Chen, Chernozhukov, and
  Hansen]{belloni2012sparse}
A.~Belloni, D.~Chen, V.~Chernozhukov, and C.~Hansen.
\newblock Sparse models and methods for optimal instruments with an application
  to eminent domain.
\newblock \emph{Econometrica}, 80\penalty0 (6):\penalty0 2369--2429, 2012.

\bibitem[Belloni et~al.(2014)Belloni, Chernozhukov, and
  Hansen]{belloni2014inference}
A.~Belloni, V.~Chernozhukov, and C.~Hansen.
\newblock Inference on treatment effects after selection among high-dimensional
  controls.
\newblock \emph{The Review of Economic Studies}, 81\penalty0 (2):\penalty0
  608--650, 2014.

\bibitem[Caruana(1995)]{caruana1995learning}
R.~Caruana.
\newblock Learning many related tasks at the same time with backpropagation.
\newblock In \emph{Advances in neural information processing systems}, pages
  657--664, 1995.

\bibitem[Chernozhukov et~al.(2016)Chernozhukov, Chetverikov, Demirer, Duflo,
  Hansen, and Newey]{chernozhukov2016double}
V.~Chernozhukov, D.~Chetverikov, M.~Demirer, E.~Duflo, C.~Hansen, and W.~K.
  Newey.
\newblock Double machine learning for treatment and causal parameters.
\newblock Technical report, cemmap working paper, 2016.

\bibitem[Chernozhukov et~al.(2018)Chernozhukov, Chetverikov, Demirer, Duflo,
  Hansen, Newey, and Robins]{chernozhukov2018double}
V.~Chernozhukov, D.~Chetverikov, M.~Demirer, E.~Duflo, C.~Hansen, W.~Newey, and
  J.~Robins.
\newblock Double/debiased machine learning for treatment and structural
  parameters, 2018.

\bibitem[Farrell et~al.(2018)Farrell, Liang, and Misra]{farrell2018deep}
M.~H. Farrell, T.~Liang, and S.~Misra.
\newblock Deep neural networks for estimation and inference: Application to
  causal effects and other semiparametric estimands.
\newblock \emph{arXiv preprint arXiv:1809.09953}, 2018.

\bibitem[Foster et~al.(2011)Foster, Taylor, and Ruberg]{foster2011subgroup}
J.~C. Foster, J.~M. Taylor, and S.~J. Ruberg.
\newblock Subgroup identification from randomized clinical trial data.
\newblock \emph{Statistics in medicine}, 30\penalty0 (24):\penalty0 2867--2880,
  2011.

\bibitem[Friedman et~al.(2001)Friedman, Hastie, and
  Tibshirani]{friedman2001elements}
J.~Friedman, T.~Hastie, and R.~Tibshirani.
\newblock \emph{The elements of statistical learning}, volume~1.
\newblock Springer series in statistics New York, 2001.

\bibitem[Friedman(2001)]{friedman2001greedy}
J.~H. Friedman.
\newblock Greedy function approximation: a gradient boosting machine.
\newblock \emph{Annals of statistics}, pages 1189--1232, 2001.

\bibitem[Goodfellow et~al.(2016)Goodfellow, Bengio, Courville, and
  Bengio]{goodfellow2016deep}
I.~Goodfellow, Y.~Bengio, A.~Courville, and Y.~Bengio.
\newblock \emph{Deep learning}, volume~1.
\newblock MIT press Cambridge, 2016.

\bibitem[He et~al.(2016)He, Zhang, Ren, and Sun]{he2016deep}
K.~He, X.~Zhang, S.~Ren, and J.~Sun.
\newblock Deep residual learning for image recognition.
\newblock In \emph{Proceedings of the IEEE conference on computer vision and
  pattern recognition}, pages 770--778, 2016.

\bibitem[Imai et~al.(2013)Imai, Ratkovic, et~al.]{imai2013estimating}
K.~Imai, M.~Ratkovic, et~al.
\newblock Estimating treatment effect heterogeneity in randomized program
  evaluation.
\newblock \emph{The Annals of Applied Statistics}, 7\penalty0 (1):\penalty0
  443--470, 2013.

\bibitem[Johansson et~al.(2016)Johansson, Shalit, and
  Sontag]{johansson2016learning}
F.~Johansson, U.~Shalit, and D.~Sontag.
\newblock Learning representations for counterfactual inference.
\newblock In \emph{International conference on machine learning}, pages
  3020--3029, 2016.

\bibitem[Kingma and Ba(2014)]{kingma2014adam}
D.~P. Kingma and J.~Ba.
\newblock Adam: A method for stochastic optimization.
\newblock \emph{arXiv preprint arXiv:1412.6980}, 2014.

\bibitem[Li et~al.(2017)Li, Ma, Le, Liu, and Liu]{li2017causal}
J.~Li, S.~Ma, T.~Le, L.~Liu, and J.~Liu.
\newblock Causal decision trees.
\newblock \emph{IEEE Transactions on Knowledge and Data Engineering},
  29\penalty0 (2):\penalty0 257--271, 2017.

\bibitem[Lu et~al.(2018)Lu, Sadiq, Feaster, and Ishwaran]{lu2018estimating}
M.~Lu, S.~Sadiq, D.~J. Feaster, and H.~Ishwaran.
\newblock Estimating individual treatment effect in observational data using
  random forest methods.
\newblock \emph{Journal of Computational and Graphical Statistics}, 27\penalty0
  (1):\penalty0 209--219, 2018.

\bibitem[Redmon et~al.(2016)Redmon, Divvala, Girshick, and Farhadi]{yolo2016}
J.~Redmon, S.~Divvala, R.~Girshick, and A.~Farhadi.
\newblock You only look once: Unified, real-time object detection.
\newblock In \emph{Proceedings of the IEEE conference on computer vision and
  pattern recognition}, pages 779--788, 2016.

\bibitem[Rosenbaum and Rubin(1983)]{rosenbaum1983central}
P.~R. Rosenbaum and D.~B. Rubin.
\newblock The central role of the propensity score in observational studies for
  causal effects.
\newblock \emph{Biometrika}, 70\penalty0 (1):\penalty0 41--55, 1983.

\bibitem[Rubin(1976)]{rubin1976multivariate}
D.~B. Rubin.
\newblock Multivariate matching methods that are equal percent bias reducing,
  i: Some examples.
\newblock \emph{Biometrics}, pages 109--120, 1976.

\bibitem[Ruder(2017)]{ruder2017overview}
S.~Ruder.
\newblock An overview of multi-task learning in deep neural networks.
\newblock \emph{arXiv preprint arXiv:1706.05098}, 2017.

\bibitem[Shalit et~al.(2017)Shalit, Johansson, and
  Sontag]{shalit2017estimating}
U.~Shalit, F.~D. Johansson, and D.~Sontag.
\newblock Estimating individual treatment effect: generalization bounds and
  algorithms.
\newblock In \emph{Proceedings of the 34th International Conference on Machine
  Learning-Volume 70}, pages 3076--3085. JMLR. org, 2017.

\bibitem[Sz{\'e}kely et~al.(2007)Sz{\'e}kely, Rizzo, Bakirov,
  et~al.]{szekely2007measuring}
G.~J. Sz{\'e}kely, M.~L. Rizzo, N.~K. Bakirov, et~al.
\newblock Measuring and testing dependence by correlation of distances.
\newblock \emph{The annals of statistics}, 35\penalty0 (6):\penalty0
  2769--2794, 2007.

\bibitem[Taddy et~al.(2016)Taddy, Gardner, Chen, and
  Draper]{taddy2016nonparametric}
M.~Taddy, M.~Gardner, L.~Chen, and D.~Draper.
\newblock A nonparametric bayesian analysis of heterogenous treatment effects
  in digital experimentation.
\newblock \emph{Journal of Business \& Economic Statistics}, 34\penalty0
  (4):\penalty0 661--672, 2016.

\bibitem[van~der Laan and Petersen(2007)]{van2007causal}
M.~J. van~der Laan and M.~L. Petersen.
\newblock Causal effect models for realistic individualized treatment and
  intention to treat rules.
\newblock \emph{The international journal of biostatistics}, 3\penalty0 (1),
  2007.

\bibitem[Van~der Laan and Rose(2011)]{van2011targeted}
M.~J. Van~der Laan and S.~Rose.
\newblock \emph{Targeted learning: causal inference for observational and
  experimental data}.
\newblock Springer Science \& Business Media, 2011.

\bibitem[Van Der~Laan and Rubin(2006)]{van2006targeted}
M.~J. Van Der~Laan and D.~Rubin.
\newblock Targeted maximum likelihood learning.
\newblock \emph{The International Journal of Biostatistics}, 2\penalty0 (1),
  2006.

\bibitem[Van~der Laan et~al.(2007)Van~der Laan, Polley, and
  Hubbard]{van2007super}
M.~J. Van~der Laan, E.~C. Polley, and A.~E. Hubbard.
\newblock Super learner.
\newblock \emph{Statistical applications in genetics and molecular biology},
  6\penalty0 (1), 2007.

\bibitem[Wager and Athey(2018)]{wager2018estimation}
S.~Wager and S.~Athey.
\newblock Estimation and inference of heterogeneous treatment effects using
  random forests.
\newblock \emph{Journal of the American Statistical Association}, 113\penalty0
  (523):\penalty0 1228--1242, 2018.

\end{thebibliography}

\end{document}